\documentclass[nofootinbib,aps,pre,twocolumn,showpacs,superscriptaddress]{revtex4-1}

\usepackage[margin=1in]{geometry}
\usepackage{amsmath}\usepackage{amsmath} 
\usepackage{amssymb}  
\usepackage{bm}
\usepackage{graphicx}
\usepackage[super]{nth}
 \usepackage{hyperref}
 \usepackage[dvipsnames]{xcolor}
\usepackage{ulem} 
\usepackage[super]{nth}
\usepackage{algorithm}
\usepackage{algpseudocode}
\usepackage{newunicodechar}

 \graphicspath{{figures/}}

\begin{document}

\title{
Bifurcation instructed design of multistate machines
}
\date{\today}
\author{Teaya Yang} \affiliation{Laboratory of Atomic and Solid State Physics, Cornell University, Ithaca, New York 14853-2501, USA}
\author{David Hathcock} \affiliation{Laboratory of Atomic and Solid State Physics, Cornell University, Ithaca, New York 14853-2501, USA}
\author{Yuchao Chen} \affiliation{Laboratory of Atomic and Solid State Physics, Cornell University, Ithaca, New York 14853-2501, USA}
\author{Paul McEuen} \affiliation{Laboratory of Atomic and Solid State Physics, Cornell University, Ithaca, New York 14853-2501, USA and Kavli Institute at Cornell for Nanoscale Science, Cornell University, Ithaca, NY, USA}
\author{James P. Sethna} \affiliation{Laboratory of Atomic and Solid State Physics, Cornell University, Ithaca, New York 14853-2501, USA}
\author{Itai Cohen} \affiliation{Laboratory of Atomic and Solid State Physics, Cornell University, Ithaca, New York 14853-2501, USA and Kavli Institute at Cornell for Nanoscale Science, Cornell University, Ithaca, NY, USA}
\author{Itay Griniasty} \affiliation{Laboratory of Atomic and Solid State Physics, Cornell University, Ithaca, New York 14853-2501, USA}

\begin{abstract}
We propose a novel design paradigm for multi-state machines where transitions from one state to another are organized by bifurcations of multiple equilibria of the energy landscape describing the collective interactions of the machine components. This design paradigm is attractive since, near bifurcations, small variations in a few control parameters can result in large changes to the system’s state providing an emergent lever mechanism. Further, the topological configuration of transitions between states near such bifurcations ensures robust operation, making the machine less sensitive to fabrication errors and noise. To design such machines, we develop and implement a new efficient algorithm that searches for interactions between the machine components that give rise to energy landscapes with these bifurcation structures. We demonstrate a proof of concept for this approach by designing magneto elastic machines whose motions are primarily guided by their magnetic energy landscapes and show that by operating near bifurcations we can achieve multiple transition pathways between states. This proof of concept demonstration illustrates the power of this approach, which could be especially useful for soft robotics and at the microscale where typical macroscale designs are difficult to implement.
\end{abstract} 
\maketitle

	\begin{figure}
		\centering
		\includegraphics[width=.95\columnwidth]{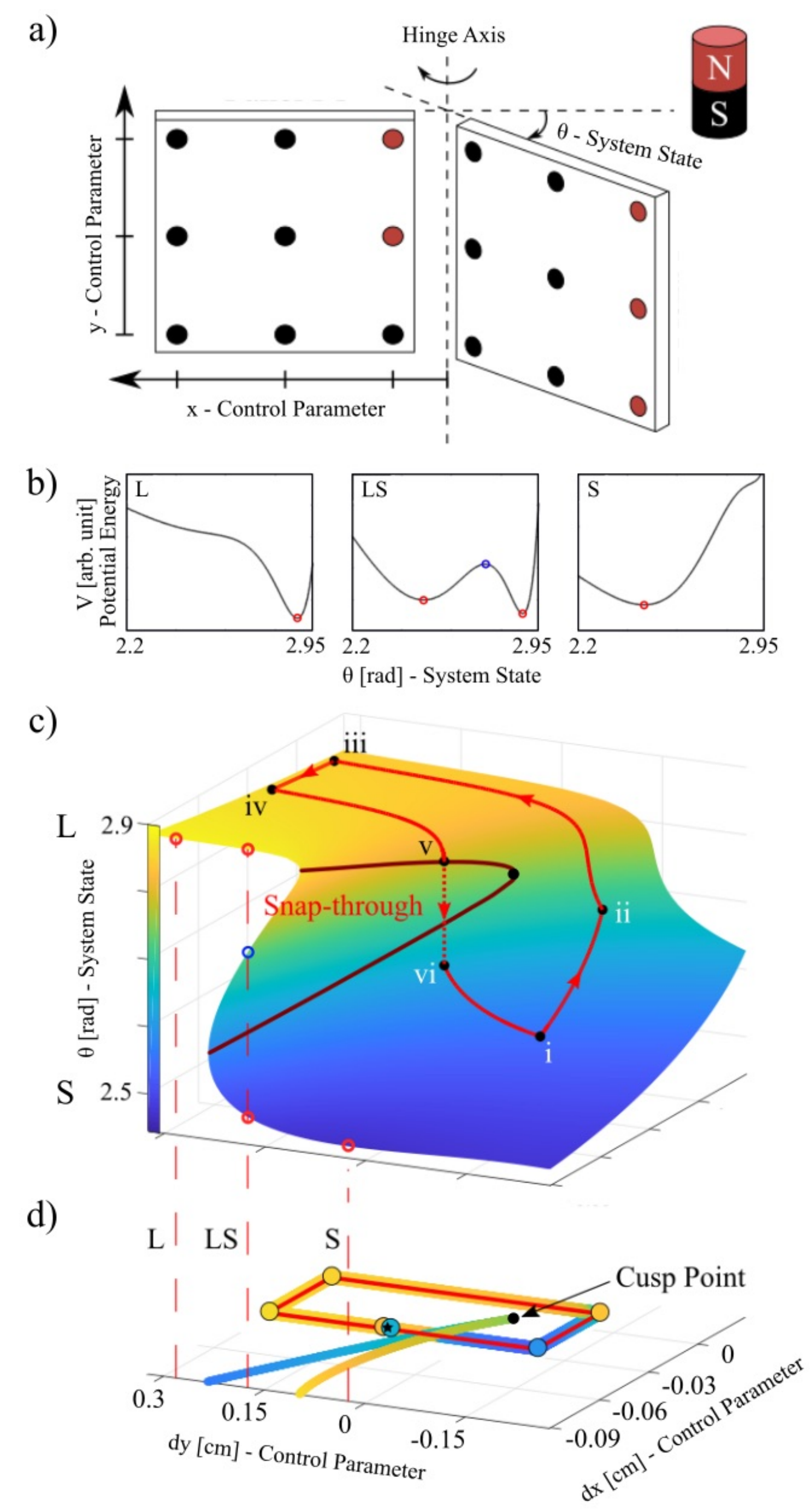}
		\caption{\textbf{Magneto-Elastic machine capable of adopting multiple configurations due to operating near a cusp bifurcation} \textbf{(a.)} \emph{ System:} Panels P1 and P2 are decorated with identical magnets. Panel P1 is actuated externally to translate in the $x$ and $y$ directions, in response Panel P2 rotates about a hinge, the dynamics are over-damped. 
			\textbf{(b.)} \emph{Magnetic potential energy landscapes:} We plot the potential for \(dx=-0.09\) and \(dy \in [0.28,0.17,0.02]\), where \(dx\) and \(dy\) are deviations from the cusp's position. Varying y we cross two saddle node bifurcations where the number of extrema of the magneto elastic landscape changes.
			\textbf{(c.)} \emph{ Equilibrium manifold:}  The system's equilibria \(\theta(dx,dy)\) are plotted as a function of the deviation of the parameters, color signifies the value of \(\theta\).
			Brown curve marks saddle node bifurcations where the number of equilibria change, and the light red curve denotes the experimental trajectory.
			\textbf{(d.)} \emph{Experimentally observed snap-through transition:}
			The system follows a parametric trajectory marked by a red curve and colored tube whose color denotes the predicted state \(\theta\), around a cusp bifurcation. The colored disks represent the experimentally measured state \(\theta\). As expected a single snap through transition at a saddle node bifurcation (curves colored according to the bifurcating state \(\theta\) and  converging at the cusp) is observed.}
		\label{Fig:landscape}
	\end{figure}

Systems composed of a large number of interacting elements such as meta-materials, elastic membranes, and proteins can exhibit emergent behaviors that arise from the collaborative interaction of the system components. Designing functionality in such systems is a formidable task that requires searches in a high dimensional parameter space of the system components and their interactions. Developing organizing principles for effectively designing such systems remains an outstanding problem in the field \cite{sigmund2013topology, goodrich2015tuning, huang2016coming, Rocks2017designing, varda2022transient, eckmann2019colloquium}. Here, we propose that designing multi-state machines around bifurcations of multiple equilibria is a powerful paradigm that can be used to systematically organize such searches.   

Bifurcations, where a single equilibrium configuration splits into multiple equilibria as a function of a control parameter is a canonical dynamical systems structure that has been used to explain various natural phenomena ranging from phase transitions \cite{pathria2016statistical} to the operation of simple machines. Examples of simple machines include Venus flytraps and hummingbird beaks that have been shown to open smoothly and then snap shut by operating about a cusp bifurcation where three equilibria converge 
\cite{venustrap,smith2011elastic}. Designing systems to operate near such bifurcations provides several advantages. Since the splitting of the equilibria has a power law dependence on the control parameters \cite{Arnold1994,Berry1977}, operating near bifurcations automatically provides a lever mechanism by which small variations in the control parameters lead to large changes in the system state \cite{overvelde2015amplifying,yinding2022snapping}. In the case of the Venus fly trap, slight changes in hydrostatic pressure can drive large motions of the trap. Similarly in hummingbirds, slight twisting of the jaw bones enables rapid closing of a wide open beak. Further, such bifurcations organize a topologically protected structure of saddle node manifolds. As such, provided that the system trajectory encircles the cusp bifurcation where the saddle node manifolds meet, the system is guaranteed to exhibit a smooth change in state followed by a snap. In the Venus fly trap and hummingbird examples, this topological protection guarantees that the opening and snapping of the trap or beak is robust against variations in the applied hydrostatic or muscle forces driving the transitions in the system state. Here, we propose that \textit{moving beyond cusp bifurcations} to design systems that \textit{operate near bifurcations of arbitrarily many equilibria} preserves the lever advantage and topological protection of cusp bifurcations.   
Such systems can be driven by only a few control parameters to undergo snapping transitions between multiple states making the design of machines near such bifurcations a powerful paradigm for organizing complex functions. 
To develop and demonstrate this paradigm, we experimentally investigate increasingly sophisticated magneto elastic machines whose function is organized by such bifurcations.
	
We start by constructing a simple magneto elastic machine consisting of a control panel that can be translated in the $x-y$ plane and a second panel that is free to rotate about a hinge connecting the two panels (Fig.~\ref{Fig:landscape}a and experimental apparatus schematic Fig.S1). The state of the system, is given by the angle $\theta$ between the panels. By decorating the panels with magnets, we are able to design a magneto elastic landscape with different numbers of minima as a function of the parameters $x$ and $y$ (Fig.~\ref{Fig:landscape}b). Transitions between these minima correspond to changes in the state of the system. To understand the various pathways for making such transitions we construct the manifold defined by the local equilibria as a function of the parameters $x$ and $y$. For this particular arrangement of magnets, we calculate (see SI) that the resulting manifold has a domain with multiple solutions delineated by saddle node bifurcation curves (Brown). These curves intersect and terminate at a cusp bifurcation beyond which there is only a single equilibrium state. By translating the control panel in the x-y plane, the system can undergo either smooth or abrupt changes in $\theta$. For example, starting the system at point (i) and moving through points (ii-v), the hinge angle increases smoothly. A further slight increase in the control parameter $y$, however, leads to an abrupt transition from a high to a low angle, corresponding to points (v) and (vi) respectively. These predictions are born out by the experiments (Fig.~\ref{Fig:landscape}d), which also show a smooth increase in $\theta$ for a pathway that encircles the cusp (i-v) and an abrupt transition in $\theta$ when crossing a saddle node curve (v-vi). In this 2D representation the system makes a transition when the color of the path (yellow) matches the color denoting the state associated with the saddle node curve (yellow). This magneto elastic mechanism is reminiscent of the cocking and snapping of a Venus flytrap or a humming bird's beak.
	
	In addition to providing a mechanism for abrupt transitions, operating near a cusp bifurcation creates a lever mechanism where small variations in the control parameters lead to large variations in the system state. This mechanism resolves the generic problem that creating large variations in the system state often requires unfeasibly large variations in the control parameters. Lever mechanisms are generic near bifurcations of equilibria since the magnitude of the transition in the system state is typically proportional to the square root of the parameter distance from the bifurcation. 
	
\begin{figure}[t]
		\centering
        \includegraphics[width=0.95\linewidth]{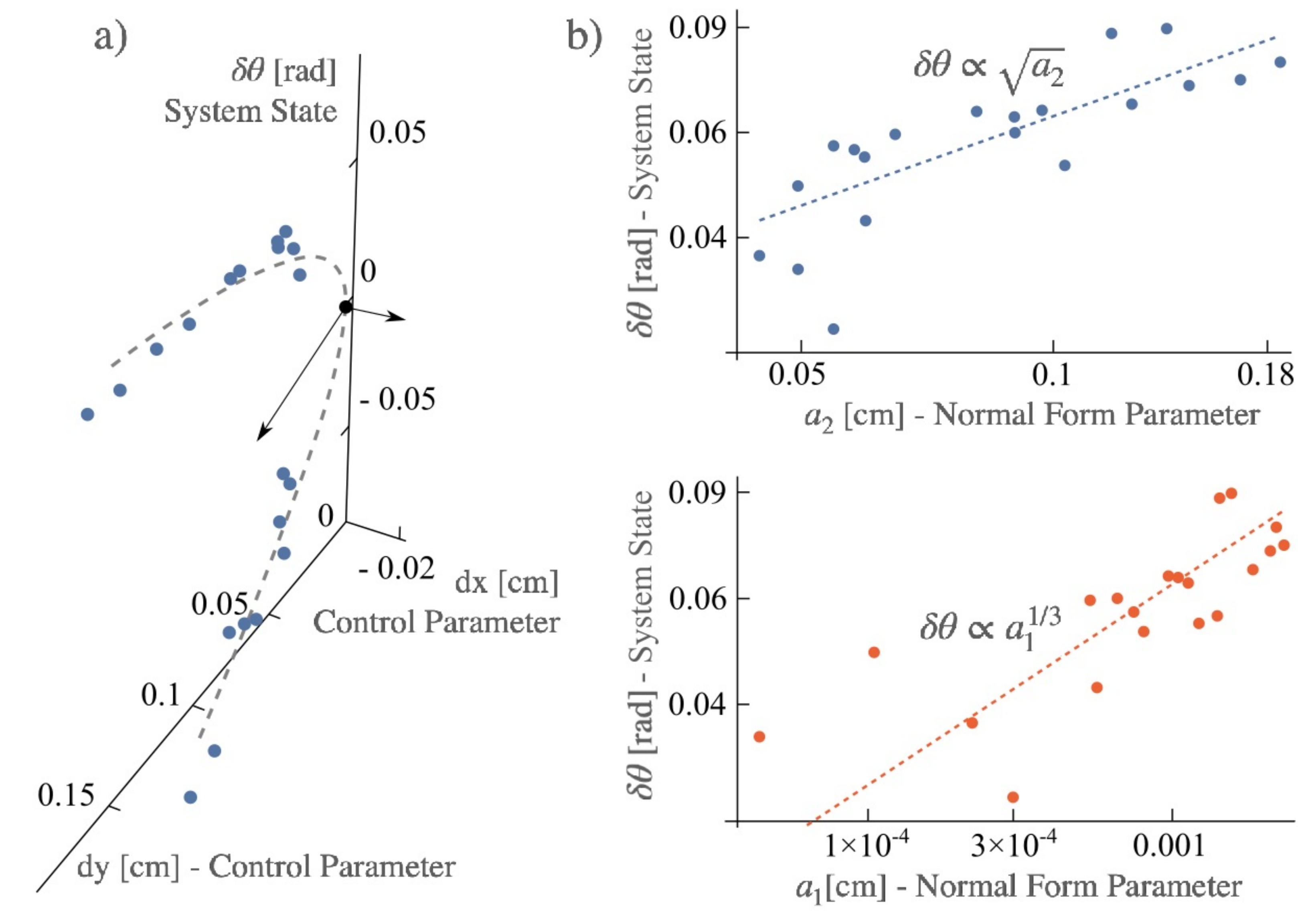}
		\caption{
			\textbf{Parametric levers}
            The change in the state of the system after a snap through transition near a cusp bifurcation scales sub linearly with the normal form parameters. This sublinear scaling leads to large variation of the state in response to small variation of the system parameters. 
            \textbf{a)} \uline{Measurements of snap through transitions near a cusp:} The blue points mark the state of the magneto elastic system of Fig.~\ref{Fig:landscape} after a snap through transition. The dashed curve is a fit of snap through transitions near a cusp bifurcation to the data, derived from the normal form potential \(\tilde{V}  = \delta\theta^4+a_2 \delta \theta ^2 + a_1\delta \theta\). The normal form parameters \(a_1\) and \(a_2\) are locally given by re-scaled rotations of \(dx\) and \(dy\), which are the deviations of the parameters away from the cusp.
	\textbf{b)} \uline{Scaling laws near a cusp:} The predicted scaling laws are demonstrated by projecting the measurements and fit onto log-log plots. Near the cusp the system response to \(a_1\) acts as a giant lever, \(\partial\delta\theta/\partial a_1\sim 50\).   
		}
		\label{Fig:Scaling}
	\end{figure}

	To characterize this lever mechanism in our experiment, we map the snapping transition curves associated with the saddle node bifurcations. Specifically, for a given value of $y$ (or $x$) we toggle $x$ ($y$) so that the system snaps back and forth, and record the values of the control parameters $x$ and $y$, and $\theta$ immediately after each transition (Fig.~\ref{Fig:Scaling}a). 
 
 To test the scaling relations, we define the normal form parameters \(a_1\) and \(a_2\) as rotations of the displacement of the parameters $x$ and $y$ from the cusp. We then fit the predicted scaling form $\Delta \theta \propto \sqrt{a_2}$ and $a_1 \propto a_2^{3/2}$ near a cusp to determine the cusp's position and the rotation of the normal form parameters. The fitted model then predicts that $\Delta \theta \propto a_1^{1/3}$ (see SI).
 Because the scaling exponents for $\Delta \theta$ are fractions of unity, small variations of the parameters along \(a_1\) and \(a_2\) lead to large variations of the system's state. For example, in our experiments the range of actuation for panel 1's position is approximately 1 cm and the range of angles accessible to panel 2 is 180$^{\circ}$ or $\pi$ radians. Near the bifurcation a translation along $a_1$ of $0.1\%$ of its range ($\sim10 \mu$ m) leads to a snap that changes $\theta$ by $\sim 5\%$ of it range ($\sim 0.1$ rad) providing a lever advantage of $\sim$50 (Fig.~\ref{Fig:Scaling}b ).   
	
	The complexity of the actions achieved by such magneto elastic mechanisms is dictated by the range and number of stable states that the system can access. This complexity can be achieved by designing the magneto elastic potentials such that the system operates near bifurcations between multiple states. 
For example, working near a hypothetical symmetric butterfly bifurcation associated with the potential $V=\theta^6+a_4\theta^4+a_2\theta^2+a_1\theta$ should enable smooth and abrupt transitions between three stable states in any order depending on the chosen trajectory for the control parameters. In Fig.~\ref{Fig:Search}a we show a cut through parameter space of the saddle node surfaces near this butterfly bifurcation. If the system starts in the S (Small) state and moves along the depicted trajectory (black arrows), it would first snap to the M (Medium) state when the system crosses the purple saddle node bifurcation and then the L (Large) state when it crosses the green curve. For the return path, however, the system would transition from the L minimum directly to the S minimum when it crosses the yellow saddle node bifurcation curve. Moreover, by working near the bifurcation, the lever mechanism should allow for transitioning between these distinct states within an accessible range of experimental control parameters. 
	
	\section*{Search algorithm for bifurcations of multiple equilibria}

To design parametric configurations corresponding to bifurcations of multiple equilibria we develop a search gradient continuation algorithm that takes advantage of their nested structure.  
Bifurcations associated with $k$ equilibria (minima plus maxima) are degenerate singularities where the first $k$ derivatives of the potential vanish. Thus they can be found iteratively by searching for singularities of the potential with increasing order, solving for one constraint at a time.  
We find that this method is especially efficient in finding experimentally realizable parametric configurations corresponding to bifurcations of multiple equilibria. Moreover, this method naturally extends to searching for bifurcations with desired properties by introducing further constraints, for example optimizing the robustness of the bifurcation's associated states to external noise.

\begin{figure}[ht!]
		\centering
		\includegraphics[width=.6\columnwidth]{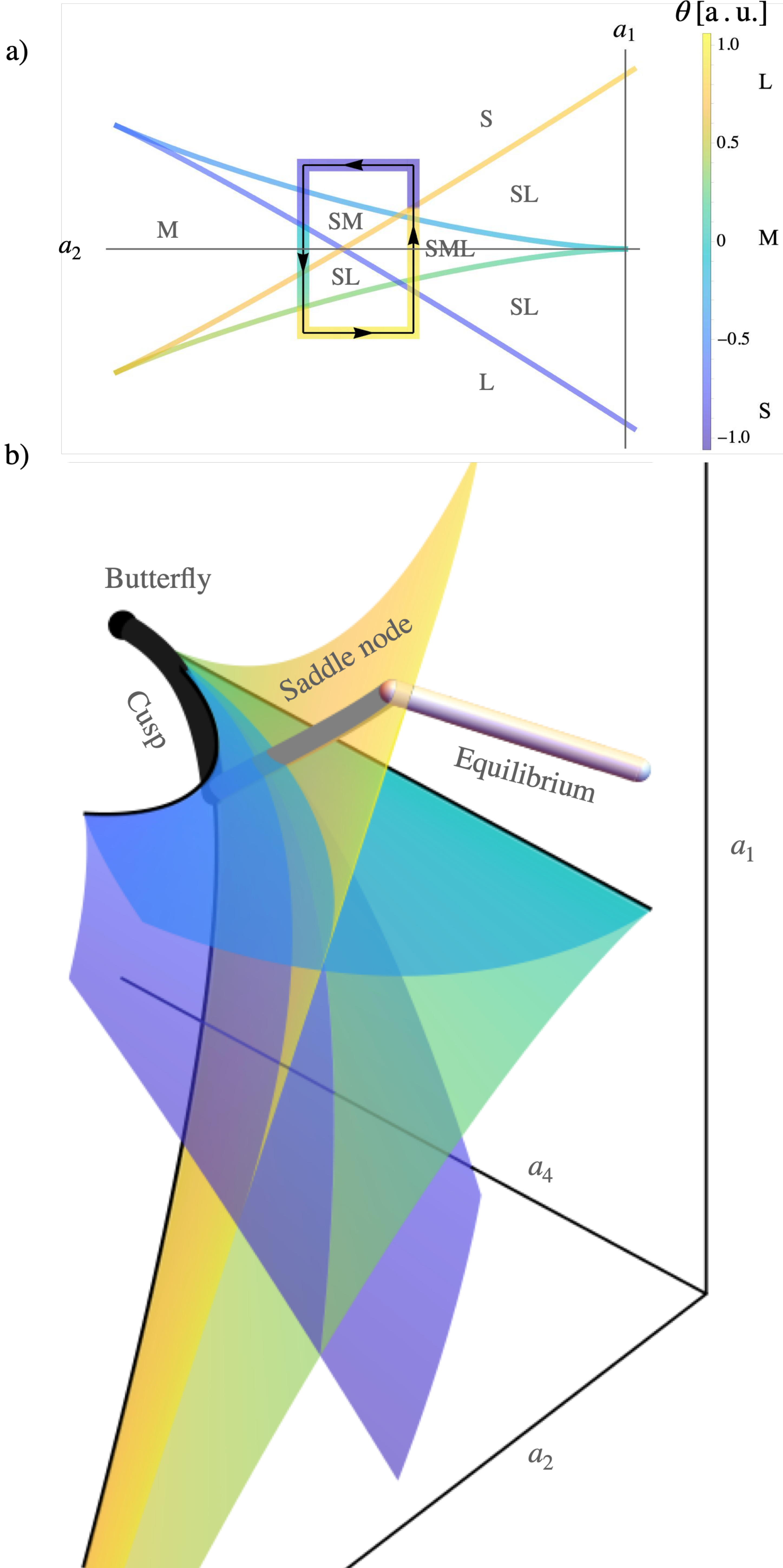}
		\caption{
		\textbf{Bifurcations of multiple equilibria. a)} \uline{Work cycle near a butterfly:} 
			A system operating near a hypothetical symmetrized butterfly bifurcation can cycle between three states. The bifurcation is associated with a potential \(V = \theta^6+a_4 \theta^4 + a_2 \theta^2 + a_1 \theta\) and three accessible states denoted by large (L), medium (M) and small (S). As the system follows the trajectory denoted by black arrows with colored background marking its state \(\theta\), it cycles between the three states snapping from S to M to L and back to S by changing \(a_2\) and \(a_1\) while \(a_4=0.1\). The snaps occur at  saddle node bifurcations (colored curves) whose color signifies the state \(\theta\) of the minima that is annihilated at each boundary.
		\textbf{b)} \uline{Gradient Continuation algorithm:} The search algorithm finds bifurcations of multiple equilibria by following a one dimensional curve. 
			Starting from a bifurcation of  \(k\) equilibria the algorithm searches for a bifurcation of \(k+1\) equilibria by following a curve in the augmented parameter space, tangent to the gradient of \(|V^{k+1}|\) in the \(k^{th}\) bifurcation manifold.
			We draw a search for a butterfly bifurcation in its symmetric normal form potential. The entire volume denotes the equilibrium manifold. Starting from a fixed point, the algorithm finds a saddle node bifurcation (along the white curve),
			Parameters are then varied on the saddle node surface (yellow), and cusp surface (thin lines) to respectively find a cusp bifurcation (along the gray curve) and a swallow tail bifurcation (along the black curve) near a butterfly bifurcation (black point).
			}
		\label{Fig:Search}
	\end{figure}

For ease of illustration we describe how to use this approach to find the symmetrized butterfly bifurcation described above with parameters $a_1,a_2,a_4$ and variable $\theta$. For a random combination of parameters we find an equilibrium angle where $dV/d\theta = 0$. Generically, this point is part of a smooth manifold over which this constraint holds. We then vary $a_1,a_2,a_4$ and $\theta$ within this manifold to minimize the next constraint $|d^2V/d\theta^2|$. The trajectory follows the gradient of the second constraint as closely as possible while maintaining the first constraint $dV/d\theta = 0$ until we reach a point on a saddle node surface, which is a manifold where both the first and second constraints hold.\footnote{A local minimum of \(|\partial_\theta^2 V|\) with respect to variation of all parameters, that lies on the fixed point manifold will throw the algorithm off, but this is a co-dimension \(m\) point for a system with \(m\) parameters, and so highly unlikely.}
Minimizing the third derivative within the saddle node manifold maintains the first two constraints and allows for finding a cusp bifurcation associated with two stable equilibria. Successive iterations allow for identifying bifurcations between an increasing number of equilibria and eventually the butterfly bifurcation. 
Our gradient continuation algorithm adapts standard algorithms from the dynamical systems literature \cite{Kuznetsov2004,PyDSTool,Guckenheimer} and retools them to locally follow the gradient of the unsatisfied constraint (see SI for further details). 
We depict the resulting search path in Fig.~\ref{Fig:Search}b, which highlights the fact that, independent of the number of parameters, the search algorithm follows a 1D trajectory, which is organized by the nested structure of the intermediate bifurcations. These properties enable the algorithm to find realizable bifurcations for systems with hundreds of parameters. 	
	
	\section*{Three states and the Butterfly Bifurcation}\label{Sec:Search} 
	As a proof of concept for our approach we demonstrate the construction and operation of a magneto elastic machine with 3 stable states operating near a bifurcation of multiple equilibria. The first step in designing such a machine is to implement our gradient continuation algorithm to design a magneto elastic potential with a butterfly bifurcation between three stable states. To realize a system operating near such a bifurcation where only three control parameters ($x$,$y$,$z$ positions of panel 1) are actively varied, we allowed the algorithm to also determine the $x$,$y$ positions of two of the nine magnets on panel 1.\footnote{Typically, a butterfly bifurcation requires four control parameters to navigate between all of the stable states. Here, we have identified a nonlinear mapping of the three active control parameters ($x$,$y$,$z$) onto the four dimensional space, which enables transitions between arbitrary minima.} With these seven parameters, the algorithm was able to identify multiple butterfly bifurcations that satisfied these criteria (See SI for details). 
	
\begin{figure*}[th!]
		\centering
\includegraphics[width=0.95\linewidth]{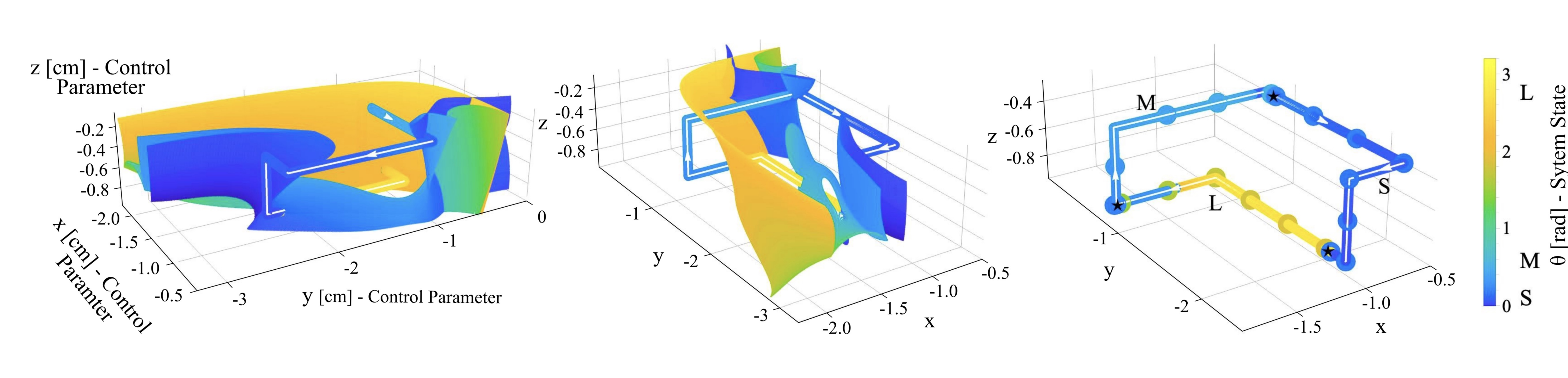}
		\caption{
			\textbf{3-state Cycle Near Butterfly Bifurcation Point} \textbf{(a.)} \uline{Theory} The saddle node surfaces of a magneto-elastic system with three active control parameters, x,y and z are plotted, their color denotes the angle \(\theta\) at which the snap occurs. The system's magnetic pattern is designed using the gradient continuation algorithm such that it operates near a butterfly bifurcation where multiple saddle node surfaces coalesce, enabling multiple snap-through transitions at the surfaces. A trajectory (colored tube with white arrows) is chosen such that the system snaps in cycles between three states Large (L), Medium (M) and Small (S) angles. The system's predicted state is denoted by the tube's color. At intersections of the trajectory with a surface where their colors match the system is predicted to snap to a new state.  \textbf{(b.)} \uline{Experimental demonstration:} The colored dots mark the experimental value of the system's state as it follows the designed trajectory. We observe three distinct transitions as predicted.
		}\label{Fig:Triple}
	\end{figure*}

	Having found an appropriate butterfly bifurcation, we use standard dynamical systems continuation algorithms\cite{guckenheimer2007cusp,Kuznetsov2004} to compute and plot the saddle node surfaces in the control parameter space ($x$,$y$,$z$) near the bifurcation (Fig.~\ref{Fig:Triple}). We find multiple distinct surfaces where the color denotes the angle $\theta$ at which the saddle node bifurcation occurs\footnote{There is further local data in the potential at a saddle node surface that can instruct the design of a trajectory. For example the sign of the third derivative of the potential signals whether the state's angle will increase or decrease as it bifurcates. Moreover, the merging of saddle node surfaces can also be delineated by plotting the cusp bifurcations. Here, we do not include this additional information for ease of viewing}. Instructed by these surfaces, we design a cyclic path through the parameter space such that the system snaps between the large, medium, and small minima. The path color at each point denotes the system state, $\theta$. As with the cusp and symmetrized butterfly bifurcations depictions in Figs.~\ref{Fig:landscape}d and \ref{Fig:Search}a, transitions occur at intersections of the path and saddle node surfaces where their colors match. We note that for the generic butterfly bifurcation, the surface structure can be quite complicated as shown by the two projections in Fig.~\ref{Fig:Triple}a,b. In contrast to the symmetrized butterfly bifurcation structure (Fig.~\ref{Fig:Scaling}a), this complicated structure necessitate using all three control parameters $x$, $y$, and $z$, to design a pathway that cycles between the three states. Importantly, despite the surface complexity the design is robust. Specifically, since the trajectory crosses surfaces, slight deviations in the control parameters should still lead to similar snaps, snap sequences, and ultimately the resulting complex actions of the entire magneto elastic machine.   
	
	Using the design parameters determined by our search algorithm, we built a magneto elastic machine similar to that depicted in Fig.~\ref{Fig:landscape}a, but with a different magnetic dipole pattern and with two of the magnets in panel 1 displaced in the panel plane (See SI). By following the theoretically predicted path, we found three snap through transitions from small to large, large to medium, and medium to small (Fig.~\ref{Fig:Triple}c and Movie S1). Two of the transitions occurred at the predicted locations, while the large to medium transition was displaced by 0.4 cm from its predicted location. In addition, we found excellent fidelity between the predicted and measured angles $\theta$ for the equilibrium states. Using the same magneto elastic machine, we also designed and demonstrated cyclical paths with two transitions (See Fig.S3 and Movie S3). Finally, when the system was taken apart and reassembled, we were able to reliably reproduce the transitions associated with the designed trajectories.  
    
	\section*{Discussion}
    
    The experimental validation of this design paradigm with a butterfly bifurcation of 5 equilibria strongly supports the conjecture that this framework could be extended to design systems performing increasingly sophisticated functions by operating near bifurcations with a growing number of equilibria. Potential energies with these increasingly rare bifurcations can be found efficiently, because the gradient continuation algorithm follows a one dimensional search path. Moreover, the associated lever mechanisms provide a design feature where the operation of the machine will likely be confined to a small parameter volume, enabling the execution of these actions by realizable machines.

Microscopic magneto-elastic machines could prove to be a useful instance of design instructed by bifurcations of multiple equilibria: An important emerging strategy for manufacturing microscopic and soft machines is fabricating them using two dimensional lithographic and printing techniques \cite{kim2012designing,Ware2015Voxelated,na2015programming,gladman16biomimetic,Miskin2020electronically}.
    Such fabrication techniques, however, restrict the implementation of compound mechanisms composed of springs, cogs, screws etc. that are used to achieve complex actions in traditional macroscale machines. 
    These lever mechanisms could be replaced with magneto elastic mechanisms with lever advantages induced by bifurcations.
    Magnetic interactions are especially well suited for this purpose since they are long ranged and not easily screened. This long range allows for global changes to the conformation in response to local actuation of system components. 
    
	Importantly, since bifurcations of multiple equilibria are notoriously sensitive to variations of parameters, there is a concern that a machine operating near such bifurcations will be very sensitive to environmental noise, such as thermal vibrations, as well as to fabrication precision. Indeed, close to a bifurcation the sensitivity of the system to variations of certain combinations of the system parameters grows exponentially as the number of associated equilibria increases. Mathematically this is captured by mapping the potential to a canonical normal form via a change of coordinates \cite[~Sec. 36.6]{Berry1977,NIST:DLMF} (see SI for derivation). Practically, however, this increased sensitivity is often blunted outside of the infinitesimal environment of the bifurcation. At a finite distance from the bifurcation the mapping to the normal form or its linearization will often cease to be valid because of other singularities of the potential or the nonlinear fall off in the potential. This non-linearity is especially pronounced in keplerian potentials such as that of magnetic interactions. Critically, the saddle node manifolds coalescing at the bifurcation are generically preserved outside this radius of convergence as they are topologically protected and can only annihilate at a cusp or a bifurcation of more equilibria. Thus, operating a machine near a bifurcation of multiple equilibria, but at a finite distance from it, allows the design of trajectories that take advantage of the multiple saddle node transitions associated with it, and their lever advantages, while avoiding the local exponential sensitivity.

    Similarly, the sensitivity of a system designed near a bifurcation of multiple equilibria to external noise grows exponentially with the number of associated states. This growth in sensitivity arises from the decrease in the potential barriers between adjacent states. For example in a potential with \(k\) equilibria where all the potential barriers are of equal height, and the minima are equally deep (which is proportional to a Chebyshev polynomial of the first kind of order \(k+1\)) the barrier heights decay as \(2^{-k}\).
    This sensitivity seems prohibitive as we imagine implementing this design principle to create systems cycling between multiple states. 
Despite this increased sensitivity, however, we estimate that the strength of magnetic interactions assures that magneto elastic systems are robust to thermal noise at the microscale. Specifically, in magneto elastic systems the potential is proportional to the dipole-dipole interaction strength \(\mu_0\mu^2 L^6/R^3\) of two magnets with magnetic dipole densities \(\mu\) panel size \(L\) and typical distance between dipoles \(R\). Thermal noise is then comparable to the magneto elastic potential barrier height when the number of equilibria \(k\sim \log_2\left(\frac{\mu_0\mu^2 L^3/(R/L)^3}{ k_b T}\right)\). The magnetic dipole densities \(\mu\) are of order \(10^6 A/m\) at the microscale \cite{cui2019nanomagnetic}. The smallest two state door (equivalent to the device in Fig.~\ref{Fig:landscape}a) that is robust to thermal noise is then \(\sim.1\mu m\) in size, approaching the size limit of 30nm for fabricating stable magnetic domains \cite{Ran2019magnetic}. Conversely, a 100 $\mu m$ machine will become sensitive to thermal noise near a bifurcation of \(\sim 40\) equilibria, that is 20 distinct states compressed in a span of \(100\) degrees.

Finally, the designs that we have implemented in this paper assume operation in a low Reynolds number regime where inertia can be neglected. In the macroscale implementation this was achieved by attaching a damping panel immersed in a solution of glycerol. We expect our designs to work even better as these machines are implemented at smaller scales since the importance of inertia drops quadratically with the system size. Operation of a 100 $\mu m$ scale machine in water, for example, would enable the system to be in the low Re regime while operating at rates that are 1000 fold faster than those in the macroscale experiment. 
	
	\section*{Conclusions}\label{sec:Discussion}
   We have shown that the operation of multi-parameter machines near bifurcations of multiple equilibria allows them to efficiently and robustly cycle between multiple conformation. Moreover, we developed a generic step-by-step framework to design and implement systems that operate near such bifurcations. Specifically, we: 1) created a search algorithm that optimizes over fabrication and other system parameters to enable operation near such bifurcations; 2) mapped the manifold of saddle node bifurcations to determine a useful trajectory for the machine operation and; 3) demonstrated the robustness of this approach by constructing and operating a magneto elastic machine that can cycle and robustly snap between multiple distinct configurations in response to small variations of a few control parameters. Importantly, this design approach and step-by-step implementation is generic and could be applied to many complex systems with multiple interacting components ranging from artificial proteins, where the interactions are electrostatic, to neural networks (both biological and synthetic) where the interactions are governed by network topology.

Cycling between transitions in mechanical implementations of such systems can generate work or locomotion. If the system is over-damped, as is often the case in microscopic systems operating in fluids, work and locomotion can be achieved by coupling the system to mechanisms that break time reversal symmetry. These mechanisms include ratchets or cilia-like flexible rods \cite{lauga2011life}. 
In the case of the magneto elastic hinge described here, time reversal symmetry is broken by combining the smooth translations of the control panel with abrupt transitions in the state of the dynamic panel. In systems where the control variable is not a mechanical parameter time reversal symmetry can be broken by using the angle as an effective dynamical variable governing a system with multiple degrees of freedom such as is often used to parameterize robot locomotion.

    More broadly, it is interesting to consider the extension of our work to systems with a larger number of dynamical variables ($\theta_1, \theta_2, \ldots$). Here, we envision that by working near bifurcations of multiple variables (e.g. elliptic umbilic bifurcations) one could organize snaps between states separated along multiple variables. Such designs require extending our search algorithm to multiple variables while maintaining its low dimensional search path.   
Alternatively, one could design mechanisms based on multiple local bifurcations that are weakly coupled across the machine. For example, one bifurcation of $n$ states could be used to control $\theta_1$ while a second bifurcation of $m$ states organizes the dynamics of the variable $\theta_2$. By weakly coupling the panels, and hence the variables $\theta_1$ and $\theta_2$, the machine can transform between $n \times m$ states in a coordinated fashion. Indeed this approach is already being implemented for bifurcations with two states \cite{coulais2018multistep,bense2021pathways,shohat2022memory}. Increasing the number of states associated with each variable would enable a similarly rich landscape for machine design with far fewer mechanical elements or panels.  

    Finally, it is interesting to consider whether this design paradigm can be used to understand natural systems beyond the Venus fly trap and hummingbird beak. For example, molecular machines such as proteins often transition between different configurations. It is interesting to consider whether such transitions can be thought of as snaps organized by bifurcations of many states \cite{huang2016coming,eckmann2019colloquium}. As another example, bifurcation theory has been implemented to identify and explain epigenetic dynamics of cell differentiation \cite{Rand2021gene,marco2014scuba, setty2016wishbone}. These approaches often focus on consecutive 2-state bifurcations. The results presented here however, suggest that a comparably simple evolutionary pathway could entail development of multi-state bifurcations. Such a structures could allow the addition of new states while maintaining the existing configuration through an evolutionary process, similar to the path taken by the gradient continuation algorithm.
    
    \section{Materials and methods:}
    		\subsection*{Construction of experimental hinge system}
        Panel P1 is constrained to a set of linear translation stages that allow its position to be adjusted manually to any $x$ or $y$ coordinates near the cusp. For experiments near the butterfly bifurcation point, an extra translation stage is attached to Panel P1 to allow adjustment of its $z$ coordinate. Panel P2 is attached to an OVA friction-less thrust air bushing with a 13mm shaft. The air bushing is attached to a fixed metal housing to limit Panel P2 to its rotational degree of freedom. A T-shaped paddle is attached to the bottom of the shaft and immersed in glycerol to introduce damping to the system. Additionally, we position a Basler Ace aca3088-57um area scan camera above the center of the air bushing to take top-view images of the air bushing which are then used to calculate the angle response of Panel P2 to high precision. 
        
        \subsection{Panels for experiments near cusp point}
        Each magnetic panel is constructed using two 1/16 in thick laser-cut acrylic pieces and nine grade N48 neodymium magnets of diameter 1/16 in and height 1/8 in. Magnets are arranged in a 3-by-3 square lattice with lattice constant of 2.5 cm.
        
        \subsection{Panels for experiments near butterfly point}
        Each magnetic panel is constructed using two 1/16 in thick laser-cut acrylic pieces and nine grade N48 neodymium magnets of diameter 1/8 in and height 1/8 in. Magnets are arranged in a 3-by-3 square lattice with lattice constant of 2.5 cm. In panel P1 the \(x,y\) position of two of the magnets is displaced according to the design determined by the search algorithm. The two magnets whose position is offset are the magnet in the bottom row on the right column, whose offsets are $dx1 = 1.418 \text{cm}$, $dy1 = -0.273 \text{cm}$, and the magnet in the middle row on the left column, with offsets $dx2 = -0.826 \text{cm}$, $dy2 = -0.986 \text{cm}$. A technical drawing illustrating the panels used for the butterfly experiment is included in the SI.

		\begin{figure}[h]
			\centering
			\includegraphics[width=0.6\linewidth]{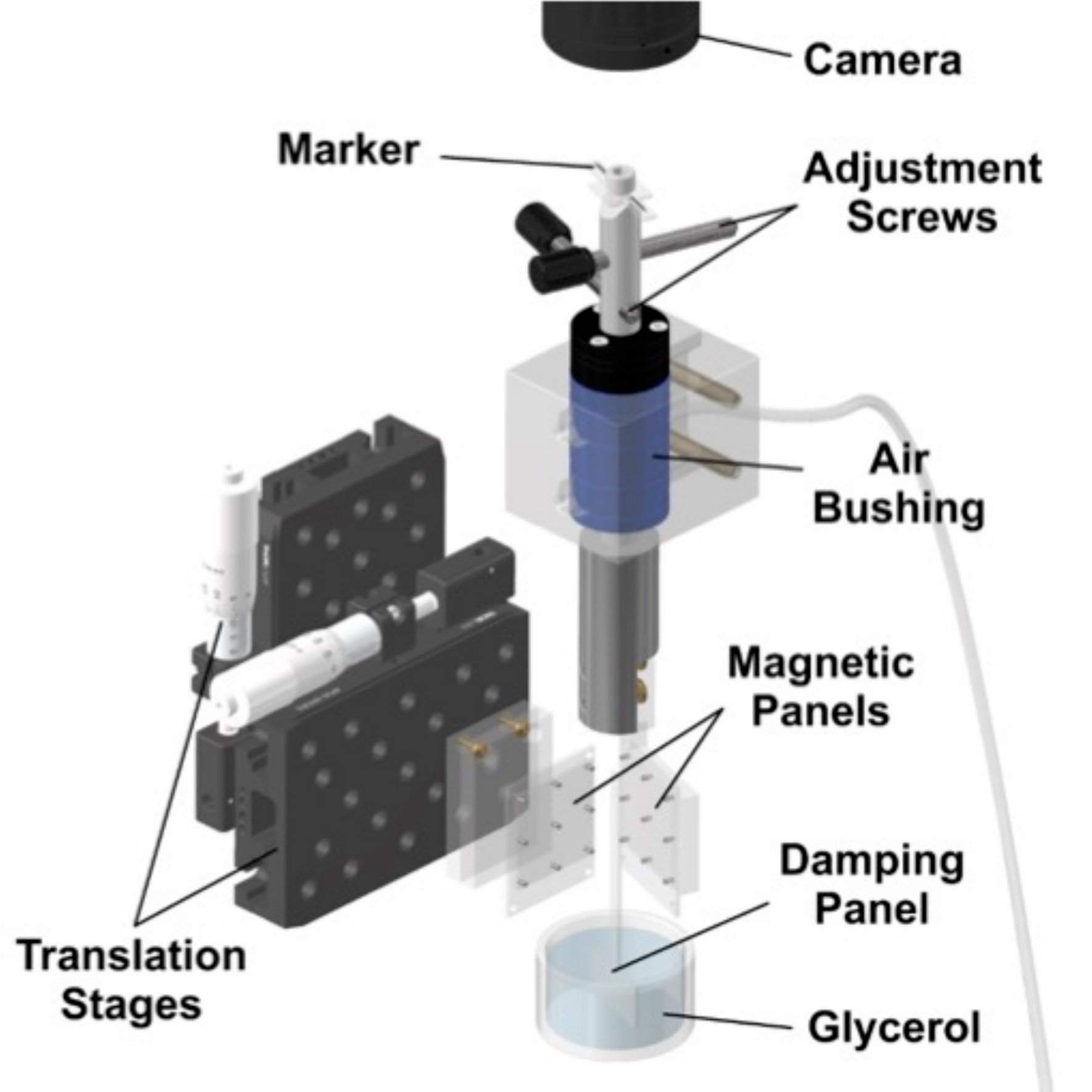}
			\caption{\textbf{Experimental Setup} Sketch of the experimental system used for demonstration of cycles and angle measurements. Panel P1 is attached to a set of translation stages which allows us to implement the spatial control parameters in all experiments. Panel P2 is attached to an air bushing that is fixed in space. An attachment submerged in glycerol is added to the base of Panel P2 to introduce damping to the system. }\label{Fig:Setup}
		\end{figure}
		
		\subsection*{Angle measurements}
            A marker is attached to the top of the air bushing, and a camera records the location of the marker during the experiment. At each given time, the measured angle is the determined by three points: current marker location, location of the center of rotation, and marker location at $\theta = 0$. We calibrate the system by recording the location of the pixel at $\theta = 0$ and several other distinct angles. The pixel location corresponding to the center of rotation is obtained using a fitted circle through the calibration data points. The resulting angle is then deduced from the three measured points. This data collection process is conducted in MATLAB.

\textbf{Acknowledgments }
We thank Michael Brenner, Chrisy Xiyu Du, Yan Yang, Robert Distasio, and John Guckenheimer for inspiring discussions. This work was financially supported primarily by NSF Grant DMREF-89228, NSF Grant EFRI-1935252, NSF Grant CBET-2010118, Cornell Center for Materials Research DMR-1719875, and  
 by Air Force Office of Scientific Research Grant MURI: FA9550-16-1-0031. I.G was also supported by the Cornell Laboratory of Atomic and Solid State Physics. D.H was supported by an NSF Graduate Research Fellowship Grant No. DGE-2139899.


{
}

\pagebreak
\widetext
\begin{center}
	\textbf{\large Supplemental material - Bifurcation instructed design of multistate machines}
\end{center}
\setcounter{equation}{0}
\setcounter{figure}{0}
\setcounter{table}{0}
\setcounter{page}{1}
\makeatletter
\renewcommand{\theequation}{S\arabic{equation}}
\renewcommand{\thefigure}{S\arabic{figure}}
\renewcommand{\bibnumfmt}[1]{[S#1]}
\renewcommand{\citenumfont}[1]{S#1}

\section{Calculation of the potential energy landscape}
To model the dynamics of our experimental hinge system, we compute the potential energy landscape arising from the dipole-dipole interactions between the magnets embedded in each panel. The magnets used in our experiments are well approximated by perfect dipoles. Therefore, the potential energy for the system is a sum of dipole-dipole interaction energies
\begin{equation}\label{dipolePotential}
V = -\sum_{i \in P1} \sum_{j \in P2}\frac{\mu_0 m^2}{4 \pi |\mathbf{r}_{ij}|^3} \left[ 3 (\mathbf{\hat m}_i \cdot \mathbf{\hat r}_{ij})(\mathbf{\hat m}_j \cdot \mathbf{\hat r}_{ij})-\mathbf{\hat m}_i \cdot \mathbf{\hat m}_j  \right],
\end{equation}
where $\mu_0$ is the vacuum permeability, $m$ is the dipole strength (identical for all magnets), $\mathbf{m}_i$ is the orientation of dipole $i$, and $\mathbf{r}_{ij}$ is the distance between magnets $i$ and $j$. Note that the interaction energy for dipoles in the same panel is constant, so we can restrict the sum to pairs of dipoles in different panels.

To derive the $\theta$ dependence of the energy landscape, we must write the dipole orientations and positions in terms of our control parameters $x$, $y$, and $z$ and the dynamical variable $\theta$. The dipoles on P1 are always oriented in the $z$-direction, while the dipoles on the rotating panel P2 have orientation that changes with $\theta$:
\begin{equation}\label{dipoleOrientations}
\begin{split}
\mathbf{\hat m}_i &= \delta_i \mathbf{ \hat z}\\
\mathbf{\hat m}_j &= \delta_j \{ \sin\theta\,,\, 0, \, -\cos \theta\},
\end{split}
\end{equation}
where  $\delta_i = \pm 1$ is the orientation of magnet $i$ with respect to panel P1 (similar for $\delta_j$). The positions of individual dipoles are given by
\begin{equation}\label{dipolePositions}
\begin{split}
\mathbf{r}_i &= \{x_i, y_i, 0\} + \{x, y, z\} \\
\mathbf{r}_j &= R_\theta \{x_j, y_j, 0\},
\end{split}
\end{equation}
leading to interdipole distance $\mathbf{r}_{ij} = \mathbf{r}_i- \mathbf{r}_j$. Here $x_i$ and $y_i$ are the $x-y$ positions of dipole $i$ in panel P1 (similar for $x_j$, $y_j$), $x$, $y$, and $z$ are the coordinates of the control panel, and $R_\theta$ is the rotation matrix corresponding to a rotation by angle $\theta$ about the $y$-axis.

Together Eqs.~(\ref{dipolePotential}-\ref{dipolePositions}) give the potential energy in terms of the hinge angle $\theta$, our control parameters $x$, $y$, and $z$, and design parameters $x_i$, $y_i$, $\delta_i$, $x_j$, $y_j$, and $\delta_j$. Since the hinge experiment is heavily damped, $\theta$ follows gradient dynamics $\dot \theta = \partial_\theta V$ and the stable equilibrium angles are given by the local minima of the potential landscape $V$.

\section{Cusp experiments}

In the cusp experiments, Panel P1's x and y positions are measured as displacements from their value when the panels are \(180^\circ\) open, and are aligned along z and y such that the panel's backs and bottoms are parallel. The magnets closest to the hinge axis are removed from it by 0.75cm on  both panels. The back of the cylindrical magnets are aligned with the panel's backs. The damping paddle used in the cusp experiments have dimensions 1.5cm by 3.0cm.

\subsection{Experimental estimation of the cusp point}
We estimate the location of the cusp point as the bifurcation of the two measured saddle node curves.
We map the saddle node curve by toggling $x$ ($y$), for a given value of $y$ (or $x$), so that the system snaps back and forth, and record the values of the control parameters $x$ and $y$, and $\theta$ immediately after each transition (Fig.~\ref{Fig:SIcusp}). Moreover, to verify the position of the cusp we record the angle $\theta$ of the system before and after snapping, and observe that the change in angle upon snapping disappears at the cusp point.

Finally, we inspect all data collected along the bifurcation curves as shown in Fig.~2a in the main text, and use a spline fit for the saddle-node bifurcations from L to S and the saddle-node bifurcations from S to L. We define the cusp point as the intersection of the two splines.

\subsection{Single snap experiment}
\begin{figure}
	\centering
	\includegraphics[width=0.95\linewidth]{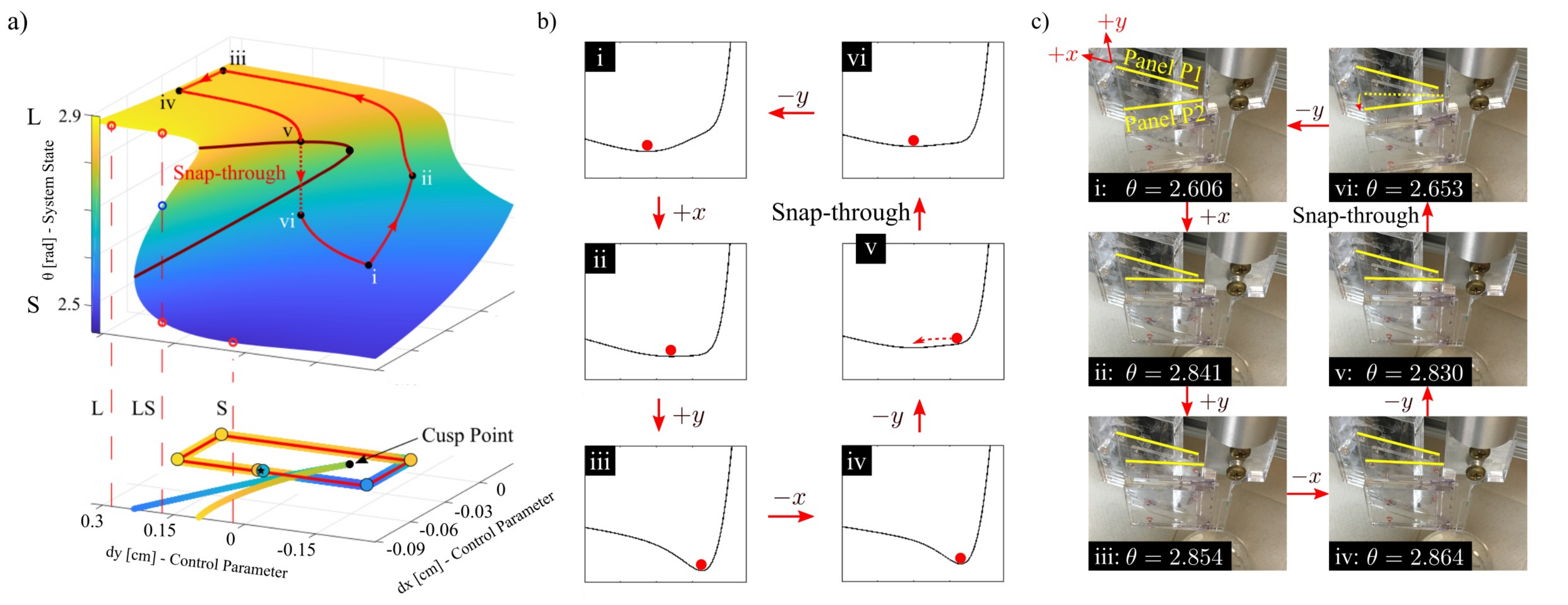}
	\caption{\textbf{Single snap-through mechanism} \textbf{(a.)} As we vary the control parameters along a loop around the cusp point as shown, we expect to see a single snap-through buckling behavior (point v to point vi) for each cycle, akin to how hummingbirds use their beak to capture prey \cite{smith2011elastic}. \textbf{(b.)} The predicted potential energy curves for points labeled from i to vi are presented. The saddle-node bifurcation occurs between v and vi as indicated by the arrow in v. \textbf{(c.)} We experimentally observe the predicted snap-through behavior. Due to experimental errors, the location of the cusp point is shifted, but we see excellent agreement between the theory and measurements after shifting the coordinates to align the theoretical and experimental cusp points. 
 }\label{Fig:Single}
\end{figure}

The mangeto elastic potential calculated for the experiment predicts a cusp at a slightly removed parametric position. The discrepency between the experimentally measured and theoretically predicted cusps could be due to fabrication errors.
To effectively compare theory and experiment in this section only, we parameterize the system as a function of its displacement from the cusp for both theory and experiment using using $dx$ and $dy$. 
We then follow the predicted path by controlling panel P1's x and y positions using the translation stages. We begin the experiment by letting the system maintain its equilibrium at the initial dx, dy position. We then change the position of Panel P1 at a slow and steady rate. Angle measurements are recorded at various locations in the loop as shown in Fig.~\ref{Fig:Single}(a) (see also Fig. 1 in the main text), and the change in position is paused once the transition happens at point vi in order to let the system settle down and obtain an accurate angle measurement. We confirm that the system returns to its original state once we return to the starting dx, dy position.

\subsection{Scaling experiment}

\begin{figure}
	\centering
	\includegraphics[width=0.95\linewidth]{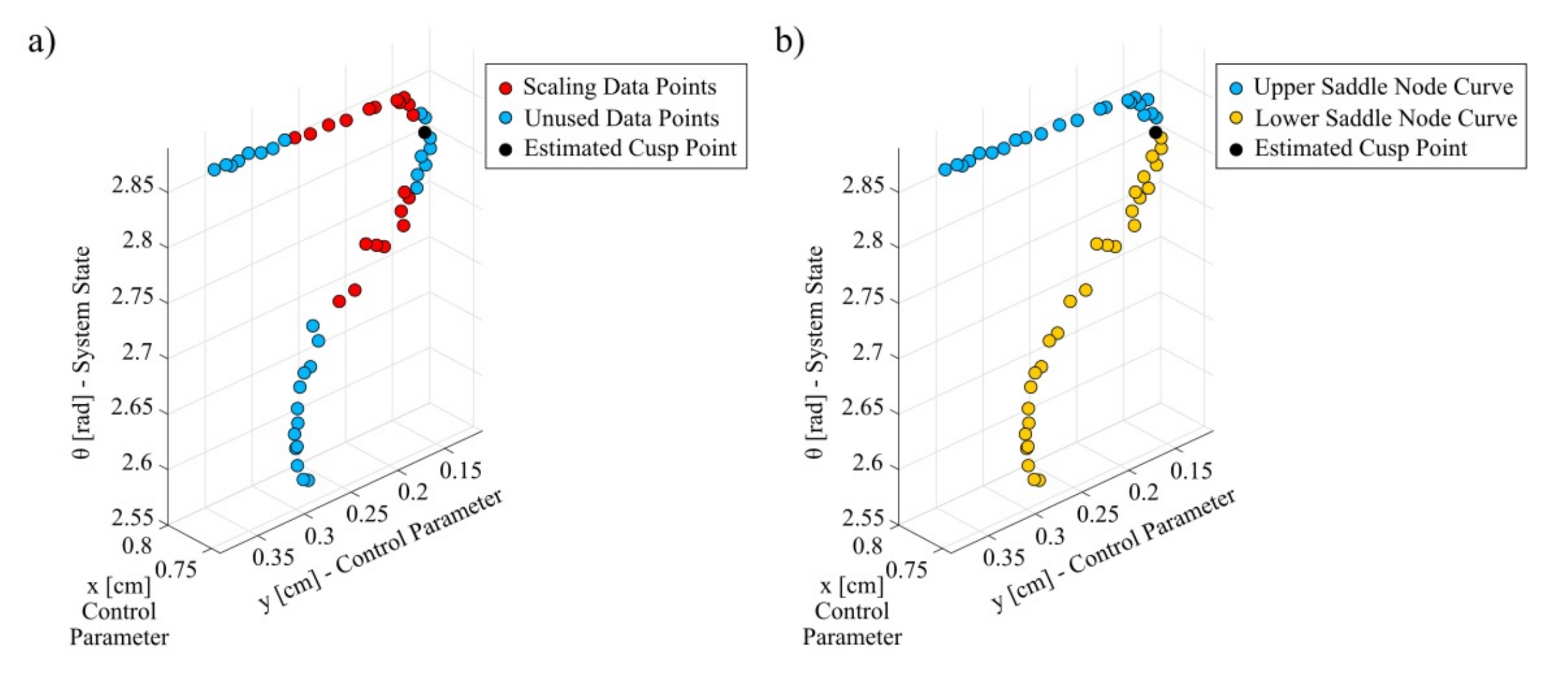}
	\caption{Snap Through transitions near a cusp. These plots show the equilibrium angle recorded in experiments following a snap-through transition. The corresponding $(x,y)$ denote the values of the control parameters at which the snap-through occurred. \textbf{(a.)} Highlights the the data points used to fit the cusp scaling. We exclude data far from the cusp, where higher order terms in the normal form are non-negligible, and close to the cusp, where measurement and fabrication error are comparable to the distance from the cusp. \textbf{(b.)} Highlights the data corresponding to the upper and lower saddle-node curves.}\label{Fig:SIcusp}
\end{figure}

To fit the scaling relations, we use the the same section of the data set used for determining the location of the experimental cusp point. We neglect data in the nonlinear region of the saddle-node curves far away from the cusp point, as well as data too close to the cusp point, where the errors due to measurement noise are comparable to the distance to the cusp. The data points used for the scaling relations are highlighted in Fig.~\ref{Fig:SIcusp}(a). The state parameter values used in the scaling analysis correspond to the angle measurements obtained at the points right after the snap through transitions.

	\section{One dimensional bifurcations of equilibria: Normal form and scaling}
    The ability to design magneto elastic machines and control parameter pathways that robustly lead to complex actions corroborates the validity of a new design paradigm: operation near bifurcations of multiple equilibria.
    The demonstrated trajectories take advantage of the structure of available dynamics near bifurcations of equilibira.
    These bifurcations are the loci of multiple distinct coalescing saddle node manifolds, as illustrated for the idealized symmetric butterfly bifurcation (Fig~3b in the main text).
    By weaving a trajectory that crosses and avoids chosen saddle node bifurcations we design a pathway that leads to complex actions.
    The system then cycles through multiple states via small variations of the control parameters, taking advantage of the multiple accessible lever mechanisms associated with these saddle node surfaces. The sensitivity of the realized design increases as the number of equilibria associated with the bifurcation grows.

	Butterfly, cusp and saddle node bifurcations are the first in a series of bifurcations of equilibria in one-dimensional gradient systems. 
	More generally, in systems with a single degree of freedom \(x\), bifurcations of $k$ equilibria are points in parameter space where the first \(k\) derivatives of the potential vanish, \(\{dV/dx, d^2V/dx^2,\ldots,d^{k} V/dx^{k}\} = \vec{0}\). That is, they are equilibrium points satisfying \(k-1\) equations beyond that of mechanical equilibrium \(d V/ dx =0\) and therefore lie on a manifold of co-dimension \(k-1\) within the equilibrium manifold.     The sensitivity of a bifurcation of \(k\) equilibria to variation in its parameters can be estimated through the topological equivalence of the dynamics near it to those in a normal form potential
    	\begin{equation}
		\widetilde{V}=\varphi^{k+1}+\sum_{i=1}^{k-1}a_i \varphi^i,
		\label{eq:cuspoid}
	\end{equation}
  where the variable \(\varphi(\theta)\) and normal form parameters \(a_i(p)\) are coordinate transformations of the angle $\theta$ and parameters $p$ respectively. The normal form describes the unfolding of the Taylor expansion of the potential at the bifurcation \(V\sim x^{k+1}\) by variations of the parameters \cite{Kuznetsov2004,Guckenheimer,bruce1992curves}. The unfolded normal form potential demonstrates that the parameteric environment of a codimension \(k\) bifurcation includes domains with \(1\) to \(\lceil(k+1)/2\rceil\) minima delineated by \(k\) saddle-node manifolds which coalesce at the bifurcation.
	Moreover, it implies scaling relations between the variation in the system's state upon a snap through transition induced by crossing a saddle node bifurcation associated with a codimension \(k-1\) bifurcation and the variation of a normal form parameter that causes the snap: 
 
	\begin{equation}
		\delta \varphi \propto a_m^{1/(k-m+1)}, \quad m < k.
	\end{equation}
	Heuristically the scaling can be derived from the normal form by noting that near the bifurcation the \(k^{\text{th}}\) derivative of the potential must still vanish, and so \(\delta\varphi^2\sim a_k\). Similarly the next \(k-1\) derivatives must progressively vanish, setting the scaling of \(a_m\). An explicit proof is given in \cite{Berry1977} and summarized in \cite[Sec. 36.6]{NIST:DLMF}.
 These scaling relations carry over to the original variable and parameters near the bifurcation where the maps \(\varphi(x)\) and \(a_m(p)\) are approximately linear.
	Indeed, the scaling relations we experimentally observed near a the cusp bifurcations are those of the systems state with the normal form parameters near a bifurcation of three equilibria, i.e., a cusp \cite{Berry1977,NIST:DLMF}.

 These scaling relations imply that the sensitivity of the system to variations of parameters grows exponentially with the number of associated equilibria. A system designed near a bifurcation of \(k\) equilibria can toggle its state between order unity separated states, \(\delta\varphi\sim 1/2\), in response to variations of the linear normal form coefficient \(a_1\) of order \(1/2^k\). That is, both the potential lever advantage and the sensitivity to noise in the parameters grow as the number of associated equilibria grows.
    However, the parametric domain in which the mapping to the normal form is linear is often very small. 
    The nonlinearity of the mapping often blunts the sensitivity of the response.
    Thus, the increased lever advantage near bifurcations of multiple equilibria is often not experimentally accessible. Conversely the system is not so sensitive to parametric noise when operated at a small parametric distance from the bifurcation about which it is designed, as demonstrated by the reproducibility of the experimental three state system, which was easily constructed twice.

\section{Continuation Algorithms}
To find bifurcations of multiple equilibria in the dynamics of our model system and to map out the saddle node structure in the vicinity of the high-order point, we use a series of continuation algorithms. In one dimension, a codimension \(k\) bifurcation point is defined by the vanishing of the first \(k\) derivatives of the potential: $ \partial_\theta^j V(\theta^*,\left\{\xi_i\right\})=0$ for $j=1,2, \dots, k$. These constraints define a codimenion \(k\) manifold in the space of dynamical variables and parameters \((\theta^*, \left\{\xi_i\right\})\).

\subsection{Traditional continuation}

Standard continuation algorithms compute bifurcation curves by varying a small number of parameters, and then projecting onto the bifurcation manifold \cite{Kuznetsov2004}. For example, suppose we have found a co-dimension \(k\) bifurcation. This requires the first $k$ derivatives of the potential vanish, fixing $\theta^*$ and $k-1$ parameters $\xi_1, \xi_2, \dots, \xi_{k-1}$. Varying an additional parameter $\xi_k$ produces a line emanating from our initial point $(\theta^*, \{\xi\}) = p$.
 The continuation algorithm maps out this line by (i) taking a step along the tangent vector $T_k(p)$ to the curve, which is the null-vector of the gradient of the first $k$ derivatives of the potential \(T_{k}(p)\equiv\left\{\vec{v}\in\mathbb{R}^{k+1} \,|\, \forall j\in(1,2,\ldots,k), \,
	\vec{v}\cdot \nabla_{\theta,\xi_1, \xi_2, \dots, \xi_{k}}\partial_\theta^jV=0 \right\}\) and (ii) correcting this step using a Newton-Raphson algorithm\footnote{ Newton-Raphson\((f, \Omega, p)\) \cite{recipes} searches for the roots of the functions \(f\) over the space \(\Omega\)  starting at the point \(p\).} to search perpendicular to the step for a point where the first $k$ derivatives of the potential vanish. This approach can be used to progressively search for higher order bifurcation points. For example, a fixed-point can be continued until $\partial^2 V(\theta^*,\left\{\xi_i\right\})/\partial\theta^2$ vanishes, indicating a saddle node bifurcation. Continuing the saddle-node can lead to a cusp bifurcation, which in turn might lead to a swallowtail bifurcation. In this way, progressively adding parameters and performing continuations of one-dimensional curves can lead toward high-codimension bifurcation points. 
	Once we have found a high-order bifurcation point, we use this algorithm to map out the saddle node surfaces nearby. The surfaces can in turn be used to design cycles in control parameters that cause the system to perform desired snapping transitions.

 
 The standard continuation approach, however, has limitations for  microscopic machine design. In particular, it has limited utility for finding the high-order bifurcation points near which our machine will operate. In our model system we have many free parameters, including the positions of each of the magnets embedded in the panels. Varying a given experimental parameter does not guarantee we will find the next order bifurcation point. Instead we want to vary many parameters simultaneously, which greatly improves the likelihood that a higher-order bifurcation point is contained within the search space and allows for a more efficient approach toward that point. We have developed a gradient continuation algorithm to carry out this multi-parameter search.

 	\subsection{Design algorithm: Gradient continuation}
		The gradient continuation algorithm works as  follows. Suppose we have $N$ parameters $\xi_i$ in our system, plus the degree-of-freedom $\theta$. A point, \(p\), where the first $k$ derivatives of the potential vanish belongs to a co-dimension $k$ manifold in the full $(N+1)$-dimensional augmented parameter space, composed of the equilibrium state and control parameters, \((\theta^*, \left\{\xi_i\right\})\). Starting from the point \(p\),  take a step along the gradient of the \(k+1\) derivative of the potential $\nabla_{\theta,\xi_1, \xi_2, \dots, \xi_{N}} \partial^{k+1}_\theta V$, projected onto the tangent surface to the manifold at \(p\). The tangent surface is  the null-space of the gradient of the first \(k\) derivatives of the potential\footnote{Notice that this algorithm uses all $N$ parameters $\xi_1, \dots, \xi_N$ to search for a codimension $k$ bifurcation, while the standard continuation in the previous section only used $k$ parameters $\xi_1, \dots, \xi_k$. The null-space $T_{k,N}(p)$ has dimension $(N-k+1)$.}, \(T_{k,N}(p)\equiv\left\{\vec{v}\in\mathbb{R}^{N+1} \,|\, \forall j\in(1,2,\ldots,k), \,
	\vec{v}\cdot \nabla_{\theta,\xi_1, \xi_2, \dots, \xi_{N}}\partial_\theta^jV=0 \right\}\). This procedure finds the step within the co-dimension $k$ manifold that maximizes the change in $\partial_\theta^{k+1} V$, which we need to vanish in order to find the next order bifurcation. After the step, the algorithm performs a corrective Newton-Raphson search \cite{recipes}, constrained to the hyperplane $T^\perp_{k,N}(p)$ perpendicular to the null-space, which returns to the codimension $k$ manifold on which the first $k$ derivatives of the potential vanish. As in the standard continuation, this approach is repeated to progressively find higher order bifurcation points.
	A visualization of the gradient search algorithm, applied to the potential $V = \theta^6 + a_4 \theta^4 + a_2 \theta^2 + a_1 \theta$, is shown in Fig.~3b in the main text.


\section{Butterfly experiments}

\subsection{Butterfly panels}

\begin{figure}
	\centering
	\includegraphics[width=0.8\linewidth]{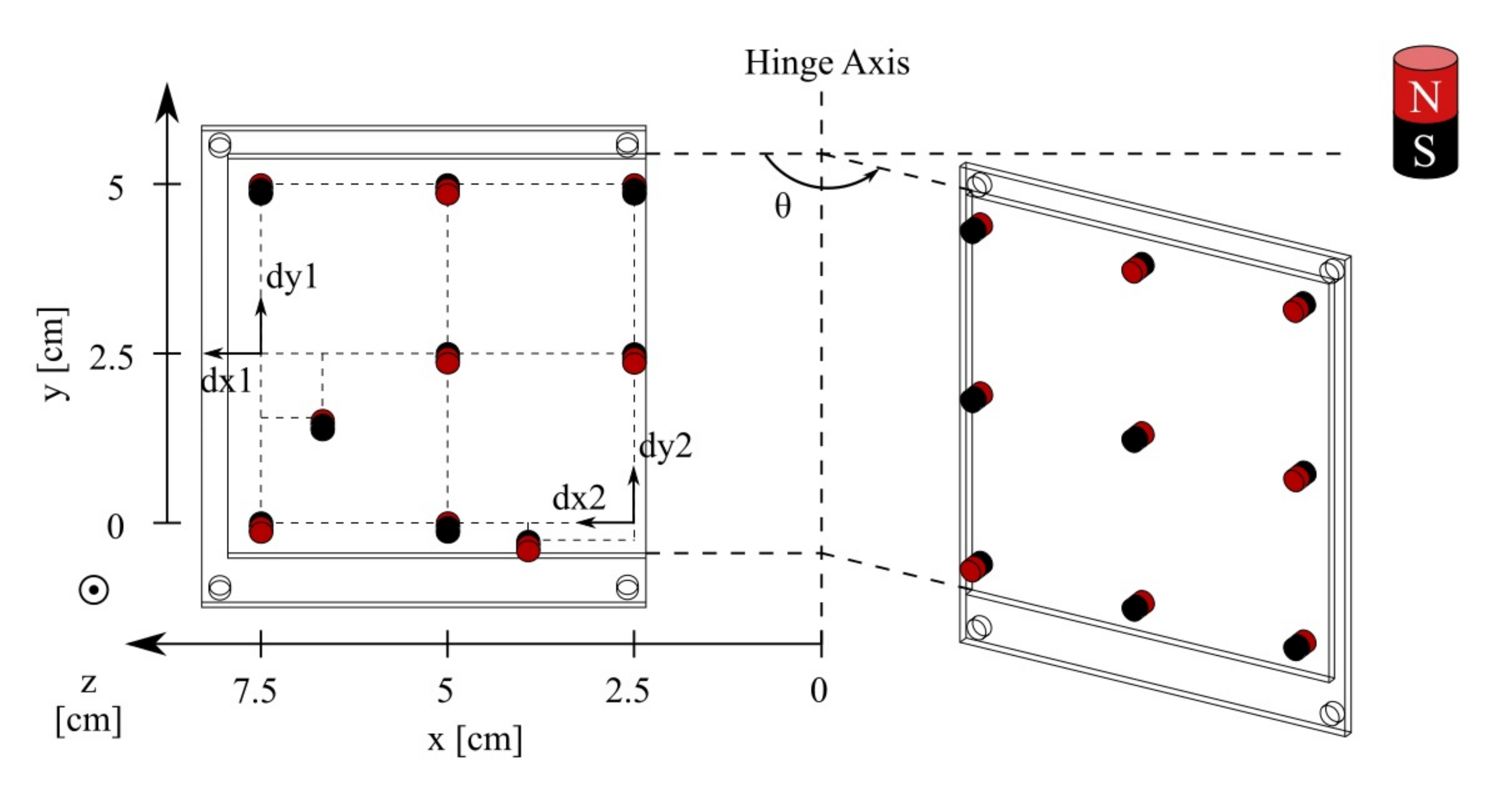}
	\caption{Butterfly panels: In the butterfly experiments, Panel P1's $x,y$ and $z$ positions are measured as displacements from their value when the panels are \(180^\circ\) open, 
the magnets closest to the hinge axis are removed from it by 2.5cm on both panels, the panels are aligned vertically and the back of the cylindrical magnets on Panel P1 are aligned with the center of the magnets on Panel P2. This small change in magnet alignment is found to reduce the discrepancy between experiment and prediction.}\label{Fig:ButterflyPanels}
\end{figure}

In the butterfly experiments, Panel P1's $x,y$ and $z$ positions are measured as displacements from their value when the panels are \(180^\circ\) open,
the magnets closest to the hinge axis are removed from it by 2.5cm on both panels, the panels are aligned vertically, and the back of the cylindrical magnets on Panel P1 are aligned with the center of the magnets on Panel P2. This small change in magnet alignment (compared to the single snap experiment) is found to reduce the discrepancy between experiment and prediction.
 An illustration for the panels is shown in Fig.~\ref{Fig:ButterflyPanels}. The damping paddle has dimensions 8.0cm by 2.5cm for the butterfly experiments. The position of the magnets on panel P1 was changed such that the system operates next to a butterfly bifurcation, as specified in the main text and in the following sections.

\subsection{Application of the continuation algorithm}

To find an experimentally feasible path and magnetic pattern, we implement the continuation algorithm by first finding a butterfly point in parameter space, then validating the resulting pattern against known experimental constraints (e.g. we require physically realizable panel angles and magnet positions). Before each search using the continuation algorithm, we first randomly generate orientations of the 18 magnetic dipoles on the two panels. We also randomly select two magnets on Panel P1 to be displaced from their lattice positions, by $(dx1, dy1)$ and $(dx2, dy2)$ respectively. 
The search algorithm is always initialized with the values \(\{\theta, dx,dy,dz,dx1,dy1,dx2,dy2\}=\{1.1\text{rad},0.5\text{cm},-0.25\text{cm},0,0,0,0,0\}\).

Next, we let the algorithm try to find a butterfly bifurcation point. If no butterfly point can be found, we repeat the initialization process and repeat the search with a new randomly generated magnetic pattern. The butterfly point corresponding to the pattern we used in our experiments is located at \sloppy\(\{\theta,dx,dy,dz,dx1,dy1,dx2,dy2\}=\{2.131 \text{rad},-0.355\text{cm},-0.304\text{cm},-0.824\text{cm},0.918\text{cm},-0.698\text{cm},-0.326\text{cm},-0.486\text{cm}\}\).

If the butterfly point is found, we investigate the potential plots at various points in parameter space near the bifurcation point. Specifically, we offset one or more of the 6 search parameters by $\pm 0.2$ and find the number of minima that exist between 0 to 180 degrees at each of these locations. The potential plots at locations with three minima are then inspected to decide the experimental feasibility of the pattern. Ideally, all three minima are at least 5 degrees apart, and the smallest minimum is at least 5 degrees (for $z=0$) to prevent the panels from touching during experiment. We also look for patterns with large triple-minima regions, for example if three visibly deep minima can be observed when at least one parameter is changed by $\pm 0.5$cm.

After an experimentally feasible pattern is discovered, we manipulate the three experimentally controllable parameters ($x$,$y$,$z$) continuously around the point with deepest triple minima and observe changes in our model of the potential landscape. The design of the control path is guided by visualization of the saddle-node surfaces mapped out using the standard continuation algorithm detailed above.  Several paths are tested in the model to obtain the desired sequence of bifurcations and to optimize various properties of the transitions (e.g. the magnitude of the snaps and depth of the minima). 

\subsection{Experiments for trajectories near a butterfly point}

We set up the experiment by laser-cutting the holes for magnets at the exact locations corresponding to the found $dx1$, $dy1$, $dx2$, $dy2$ values, which were $1.418$cm, $-0.273$cm, $-0.826$cm, and $-0.986$cm respectively. We also add a translation stage to control Panel P1's z position. We begin the experiment by following the exact coordinates provided by the theoretically designed path. In the event that a predicted transition cannot be seen using the predicted path coordinates (due to fabrication or calibration errors shifting the surface), we translate the system further from the original predicted path to determine a more robust path that may account for some shifting in coordinates due to experimental errors (for example see Fig.~4b in the main text). Once an experimental path is shown to demonstrate the predicted behavior with the desired number of state transitions, we record the locations for state transitions in experiment, and repeat the experiment while slowing down the rate of change in x,y positions near the transitions to give the system enough time to respond in the presence of large damping.  Those experiments show excellent qualitative agreement with the theoretically designed paths, although the locations at which transitions happen and the equilibrium angle of the panel are often shifted by a small amount due to experimental error.

\subsection{Additional operation mode: double-snap trajectories}
The intricate saddle-node surface structure near the butterfly bifurcation enables a variety of snapping behaviors with the same panel design, beyond the 3-state cycle presented in the main text. Here we present a second snapping sequence that was measured experimentally. 

\begin{figure}
    \centering
    \includegraphics[width=\linewidth]{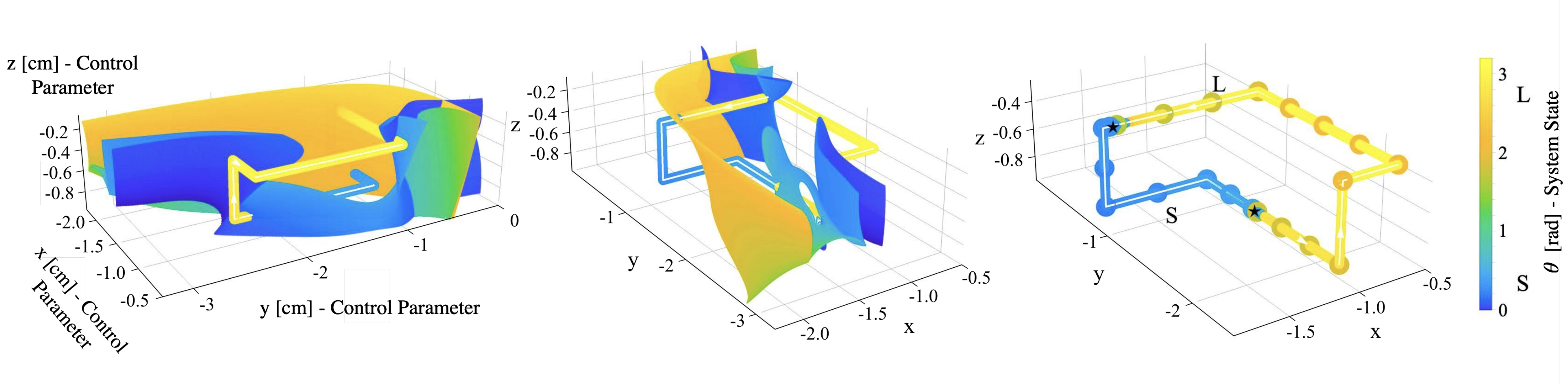}
    \caption{
    \textbf{2-state Cycle Near Butterfly Bifurcation Point} { \textbf{(a.)} \uline{Theory} The saddle node surfaces of a magneto-elastic system with three active control parameters, x,y and z are plotted, their color denotes the angle \(\theta\) at which the snap occurs. The system's magnetic pattern is designed using the gradient continuation algorithm such that it operates near a butterfly bifurcation where multiple saddle node surfaces coalesce, enabling multiple snap-through transitions at the surfaces. A trajectory (colored tube with white arrows) is chosen such that the system snaps back and forth between two states with Large (L) and Small (S) angles. This trajectory is identical to that for the 3-state cycle in Fig.~4 in the main text, but the path is traversed in the opposite direction. The system's predicted state is denoted by the tube's color. At intersections of the trajectory with a surface where their colors match the system is predicted to snap to a new state.  \textbf{(b.)} \uline{Experimental demonstration:} The colored dots mark the experimental value of the system's state as it follows the designed trajectory. We observe two distinct transitions as predicted.}
    }\label{Fig:double}
\end{figure}

By using the same trajectory in parameter space as the three-snap sequence in the main text, but traverses the path in the reverse direction, we observe a two-snap sequence between small (S) and large (L) angles. Fig.~\ref{Fig:double}a shows this trajectory together with the same saddle surfaces from Fig.~4a in the main text. The experimentally measured angles along this backward cycle are shown in Fig.~\ref{Fig:double}b (see also Movie S2). Besides a minor systematic shift in the angles of the L state, we find excellent fidelity between the predicted and measured angles. The snapping transitions occur almost exactly at the predicted locations. 

Our example trajectories demonstrate that the saddle-node structure in the vicinity of a butterfly bifurcations enables a great deal of flexibility in controlling state transitions of a mechanical system. For practical applications, further fine-tuning of the control trajectory can be used to optimize features the system's behavior (e.g., the positions of the steady states and their lifetimes in the presence of environmental noise).

\section{Generalizations: multidimensional bifurcations  and supplemental scaling behaviours}

\subsection{Stopping conditions in higher dimensions}

While our proof-of-concept experiment is limited to a hinge with a single degree of freedom (the opening angle), our approach and gradient continuation algorithm are straightforward to apply to systems with multiple degrees of freedom, e.g. a microscopic robot with multiple panels connected by elastic hinges. The cuspoidal bifurcations discussed in this paper also naturally appear in higher-dimensional gradient systems. However, the analytic criteria to classify them is somewhat more complicated: we can not simply search for points where higher order derivatives of the potential vanish. In this section we will discuss stopping criteria in higher dimensions, i.e. what quantities should we follow with the gradient continuation algorithm to search for bifurcations of increasing order?

With two or more degrees of freedom, a saddle-node bifurcation occurs when a fixed-point (stable or unstable) collides with a saddle point, resulting in mutual annihilation. This occurs when an eigenvalue of the Hessian of the potential $A_{ij} = -\partial_{\theta_i} \partial_{\theta_j} V$ crosses 0 (here $\theta_i$ are the dynamical variables). For the purposes of applying gradient continuation starting from a fixed point, it is therefore convenient to use $\det A$ as the stopping criteria, since the determinant vanishes when an eigenvalue does. 

Near a saddle-node bifurcation, the state space can be decomposed (by the Center Manifold Theorem) into (i) the invariant center manifold emanating from the fixed point along the direction of the critical eigenvector (with eigenvalue 0) and (ii) a stable/unstable manifold on which the flows exponentially grow or decay (for the purposes of machine design we generally want only stable directions). Due to the vanishing eigenvalue, the dynamics on the center manifold are nonlinear at lowest order. These dynamics can be determined perturbatively by expanding the gradient of the potential, projecting onto the center manifold and enforcing the invariance of the center manifold \cite{Kuznetsov2004}. Higher-order bifurcations occur when the center manifold expansion coefficients vanish. For example, vanishing quadratic term  indicates a cusp bifurcation, vanishing cubic term indicates a swallowtail, and so on. Thus these coefficients replace the higher-order derivatives of the potential as the stopping criteria in the gradient continuation algorithm. Below we give explicit expressions for these expansion coefficients.

Suppose we have an $n$-dimensional system $\theta \in \mathbb{R}^n$ that undergoes a saddle node bifurcation at $\theta=0$. Near this point, the dynamics can be expanded as follows,
\begin{equation}
\dot \theta = A \, \theta +F(\theta),
\end{equation}
where $A$ is the Hessian of the potential (which has a zero eigenvalue) and $F(\theta)$ collects all quadratic and higher-order terms in multilinear forms,
\begin{equation}
\begin{alignedat}{2}
F(\theta) &= \frac{1}{2} B(\theta,\theta) + \frac{1}{6}C(\theta, \theta, \theta) \, +&& \frac{1}{24}D(\theta, \theta, \theta, \theta) + \mathcal{O}(||\theta||^5)\\
& = \frac{1}{2} \sum_{i,j=1}^n \frac{\partial^2 F(\phi)}{\partial \phi_i \partial \phi_j}\bigg|_{\phi=0} \theta_i \theta_j\,+ &&\,\frac{1}{6}\sum_{i,j,k=1}^n \frac{\partial^3 F(\phi)}{\partial \phi_i \partial \phi_j \partial \phi_k}\bigg|_{\phi=0} \theta_i \theta_j \theta_k \\
& && + \frac{1}{24}\sum_{i,j,k,l=1}^n \frac{\partial^4 F(\phi)}{\partial \phi_i \partial \phi_j \partial \phi_k \partial \phi_l}\bigg|_{\phi=0} \theta_i \theta_j \theta_k \theta_l+  \mathcal{O}(||\theta||^5).
\end{alignedat}
\end{equation}
Let $\psi$ and $\varphi$ be the right and left eigenvectors corresponding to the zero eigenvalue: $A \psi = 0$ and $A^T \varphi = 0$. The projection of $\theta$ onto the center manifold $\vartheta = \varphi \cdot \theta$ has dynamics
\begin{equation}
\dot \vartheta = a_2 \vartheta^2 + a_3 \vartheta^3 + \mathcal{O}(\vartheta^4).
\end{equation}
Following Kuznetsov, we derive the coefficients up to fourth order (third order is given in Ref.~\cite{Kuznetsov2004}),
\begin{equation}
\begin{split}
a_2 &= \frac{1}{2} \varphi \cdot B(\psi, \psi)\\
a_3 & = \frac{1}{6} \varphi \cdot C(\psi, \psi, \psi) + \frac{1}{2} \varphi \cdot  B(\psi, b_2)\\
a_4 & = \frac{1}{24} \varphi \cdot D(\psi, \psi, \psi, \psi)+\frac{1}{4}\varphi \cdot C(\psi, \psi, b_2) + \frac{1}{8}\varphi \cdot B(b_2,b_2) + \frac{1}{6} \varphi \cdot B(\psi, b_3),
\end{split}
\end{equation}
where
\begin{equation}
\begin{split}
b_2 &= A^{-1}_{su} \Big (\psi [\varphi \cdot B(\psi, \psi)]-B(\psi, \psi)  \Big) \\
b_3 & = A^{-1}_{su} \Big( 
 \psi [\varphi \cdot  C(\psi, \psi, \psi) + 3 \varphi \cdot B(\psi, b_2)] + 3 b_2 [\varphi \cdot B(q, q)]  - C(\psi, \psi, \psi) -3 B(\psi, b_2)\Big)
\end{split}
\end{equation}
and $A^{-1}_{su}$ is the inverse of $A$ restricted to the stable/unstable subspace (which doesn't have zero eigenvalues). As mentioned above, vanishing $a_2$ indicates a cusp, if $a_3$ also vanishes we have a swallowtail, and if all three coefficients are zero we have a butterfly. The vectors $b_2$ and $b_3$ describe the curvature of the center manifold in the full $\theta$ space, $\theta =  q \vartheta +  b_2 \vartheta^2/2 + b_3 \vartheta^3/6$. While these bifurcations are one dimensional (they occur on the one-dimensional invariant center manifold), the curvature of the center manifold as we move further from the bifurcation point could allow snapping between states with reasonable separation in multiple dimensions. In principle, this would enable machines to carry out work cycles near a butterfly bifurcation.

\subsection{Scaling for the Thom's seven: hyperbolic and elliptic umbilics}
Beyond the quasi-one-dimensional bifurcations there are also cuspoidal bifurcations that are genuinely multidimensional. In two dimensions, for example, we have elliptic umbilic, hyperbolic umbilic, and parabolic umbilic catastrophes (these together with the four one-dimensional bifurcations saddle-node, cusp, swallowtail, and butterfly make up the Thom seven). Like the cusp and butterfly bifurcations, the unfolding of the normal form predicts and intricate saddle-surface structure describing how fixed-points and saddle-points come together and collide in the vicinity of the bifurcation point. These higher-dimensional bifurcations also obey advantageous scaling laws, relating the changes in state to the variation of control parameters. For example, the normal form potentials for the elliptic and hyperbolic umbilics are
\begin{equation}
\begin{split}
V_\text{elliptic} &= \frac{\theta_1^3}{3} -\theta_1 \theta_2^2 + a (\theta_1^2 + \theta_2^2) + b \theta_1 + c \theta_2 \\
V_\text{hyperbolic} &= \theta_1^3 + \theta_2^3 + a \theta_1 \theta_2 + b \theta_1 + c \theta_2 
\end{split}
\end{equation}
from which the follow scaling can be derived \cite{Berry1977},
\begin{equation}
\delta \theta_1,\delta \theta_2\sim a\quad b,c\sim a^2.
\end{equation}
Increasing the dimension further leads to even more cuspoidal bifurcations; these have been enumerated by Arnold using an ADE classification \cite{Arnold1994}. While the search criteria for such bifurcations is increasingly complicated, they provide a rich design space for multi-component machines.

\subsection{Reynolds number scaling}

The magnetic decorations in our experiments are arranged in each panel about a square lattice with unit separation of \(2.5\)cm.
	To explore over-damped, gradient dynamics, that are ubiquitous in microscopic mechanisms, the rotating panel is attached to a paddle moving through a  glycerol bath. The results of our experiments then hold also for smaller systems in fluid with comparable kinematic viscosity.
	If the system is smaller by a factor \(\Omega\ll 1\), the time \(\Delta t\) it takes our macroscopic over-damped system, of typical size \(L\), to traverse an angular expanse \(\Delta\theta\) is equal to the time it takes a microscopic system, of size \(\Omega L\) to traverse the same angular expanse in the same liquid. This comes about because both the viscous drag force and the magnetic force between dipoles of magnetization \(M_1\) and \(M_2\), \(F_\text{Drag}\sim L^2 \dot{\gamma}\), \( F_\text{dipole}\sim M_1 M_2/R^4\), are quadratic in the typical system sizes. For over-damped dynamics this results in a length-scale independent strain-rate, \(\dot{\gamma}\).
	The system is over-damped if its Reynolds number \(\mathrm{Re}=L^2 \dot{\gamma}/\nu\), is smaller then \(1\), where \(\nu\) is the fluid's kinematic viscosity.
	The Reynold's number of a miniaturized system is therefore smaller by a factor of \(\Omega^2\). 
	Reducing the system's size can compensate for changes in the system's composition, such as embedding it in water rather than glycerol, or the growth of magnetic dipole strength density as the system size decreases.

%

\newpage

\end{document}